
\input phyzzx
\nonstopmode
\sequentialequations
\overfullrule=0pt
\tolerance=5000
\nopubblock
\twelvepoint

\REF\rwone{K. Rajagopal, F. Wilczek {\it Nuclear Physics\/}
{\bf B399} 395 (1993).}

\REF\rwtwo{K. Rajagopal, F. Wilczek Princeton preprint
PUPT-1389, 1993 -- to appear in {\it Nuclear Physics\/} {\bf B}.}

\REF\pol{ A. Polyakov {\it Phys. Lett}. {\bf 72B} 477 (1978);
L. Susskind {\it Phys. Rev}. {\bf D20} 2610 (1979).}

\REF\sy{B. Svetitsky, L. Yaffe {\it Nuclear Physics\/} {\bf B210}
[FS56] 423 (1982).}

\REF\nmglue{J. Kogut {\it et al}. {\it Phys Rev. Lett}.
{\bf 48} 1140 (1982); {\bf 50} 393 (1983); {\bf 51} 896 (1983).}

\REF\pw{R. Pisarski, F. Wilczek {\it Phys. Rev}. {\bf D29} 338
(1984).}

\REF\bnm{G. Baker, B. Nickel, D. Meiron {\it Phys. Rev}. {\bf B17}
1365 (1978); {\it Compilation of 2-pt. and 4-pt. graphs for
continuous spin models\/} U. of Guelph report, unpublished (1977).}

\REF\pat{A. Paterson {\it Nuclear Physics\/} {\bf B190} [FS3]
188 (1981).}

\REF\gs{H. Gausterer, S. Sanlievici {\it Phys. Lett}. {\bf B209}
533 (1988).}

\REF\nmquark{Reviewed by S. Gottlieb, {\it Nucl. Phys}. {\bf B}(Physics
Supplement){\bf 20} 247 (1991).}

\REF\bm{See for example T. Newman, A. Bray, M. Moore
{\it Phys. Rev}. {\bf B49} 4514 (1990).}

\REF\txtr {N. Turok, D. Spergel {\it Phys. Rev. Lett}. {\bf 66}
3092 (1991).}

\REF\misal{ A. Anselm, M. Ryskin {\it Phys. Lett}. {\bf 226} (1991);
J.-P. Blaizot, A. Krzywicki {\it Phys. Rev}. {\bf D46} 246 (1992);
J. Bjorken, {\it Int. J. Mod. Phys}. {\bf A7} 4189 (1987),
{\it Acta Physica Polonica\/} {\bf B23} 561 (1992); K. Kowalski, C.
Taylor preprint hepph/9211282 (1992); J. Bjorken, K. Kowalski, C.
Taylor SLAC preprint SLAC-PUB-6109 (1993).}

\REF\cft{For an excellent review with extensive references see
P. Ginsparg {\it Applied Conformal Field Theory\/} in {\it Fields,
Strings, and Critical Phenomena\/} Les Houches summer school 1988,
ed. E. Brezin, J. Zinn-Justin (North Holland, Amsterdam 1990).}

\REF\cole{S. Coleman {\it Comm. Math. Phys}. {\bf 31} 259 (1973).}

\REF\merm{N. Mermin H. Wagner {\it Phys. Rev. Lett}. {\bf 17}
1133 (1966).}

\REF\hz{H. Hanson, I. Zahed, this conference.}

\line{\hfill IASSNS-HEP-93/48}
\line{\hfill July 1993}
\line{\hfill hep-ph/9308341}
\titlepage
\title{Remarks on Hot QCD\foot{ Invited talk given at
the Quark Matter '93 conference,
Borlange Sweden June 1993.}}
\author{Frank Wilczek\foot{Research supported in part by DOE
grant
DE-FG02-90ER40542}}
\vskip.2cm
\centerline{{\it School of Natural Sciences}}
\centerline{{\it Institute for Advanced Study}}
\centerline{{\it Olden Lane}}
\centerline{{\it Princeton, N.J. 08540}}
\endpage

\abstract{After briefly reviewing work I've
done recently with Krishna
Rajagopal on the nature of the equilibrium phase
transition in QCD at finite temperature and zero baryon number
and on the possibility of forming misaligned vacuum by a quench,
I discuss -- also briefly -- two new items.
These are extension of the arguments
concerning the order of the transition to
to 2+1 dimensions, relating in this case gauge to
conformal field theories, that could be explored in numerical experiments;
and the energy budget at the chiral transition and its relevance to
the nature of the transition and to quenching.}

\endpage


The bulk of the material in the talk as delivered was taken from
two papers written with Krishna Rajagopal.  Here I will
only give a telegraphic summary of those papers
[\rwone , \rwtwo ] in Sections
1 and 2.  Sections 3-4 are devoted to the new items mentioned in
the Abstract.

\chapter{Equilibrium Phase Transition}

Because of asymptotic freedom, one expects that at sufficiently
high temperature hadronic matter in thermal equilibrium will
form a weakly interacting plasma of quarks and gluons.  Such a
plasma differs qualitatively from hadronic matter
at low temperature
in two important respects.  First, of course at low temperatures
one does not observe quarks and gluons but rather color neutral
objects -- the property of confinement.  Second,
a large body of phenomenology connected with soft pion theorems
indicates that
$SU(2)\times SU(2)$
chiral symmetry suffers a large spontaneous breakdown,
in addition to being intrinsically but
relatively weakly violated
by the existence of
small but $u$ and $d$ quark masses.   It may also be useful to consider
$SU(3)\times SU(3)$ chiral symmetry, though in that case it is
unclear when it is
appropriate to consider the intrinsic violation as small.
In any case, the spontaneous portion of the breaking clearly is
not contained in the standard description of QCD as a
weakly interacting quark-gluon plasma at high temperature.

It is natural to ask whether these qualitative changes
are marked by
actual phase transitions, and if so what is
the nature of these transitions.
Is there a first order transition with discontinuities in
thermodynamic functions and release of latent heat, or a second order
transition where the primary thermodynamic functions are continuous
but not analytic -- or perhaps no true phase transition at all?  In
this connection let me remind you that the passage of ordinary neutral
gas into an ionized plasma at high temperature, which in some respects
is analogous to deconfinement, although it connects two very different
kinds of matter, is {\it not\/} marked by any singular behavior in the
thermodynamic functions.

How does one approach such questions theoretically?  A powerful method
is to construct {\it order parameters}, usually characterizing the state
of symmetry of the system, that are zero in one phase but not in the other.
Classic examples of order parameters are the magnetization or staggered
magnetization which characterize
ferromagnetic respectively antiferromagnetic
states.  These order parameters
necessarily vanish in a state with rotational symmetry, such
one expects above the Curie or Neel point.
If one has an order parameter of this kind
one can be sure that there is a singularity in passing
between the states, because an
analytic function cannot vanish in a finite range without vanishing
everywhere.  On the other hand one may have phase transitions with
no order parameter -- the thermodynamic functions on the two sides
of the transition temperature may just be different functions.  This
is usually the case for liquid-gas transitions.  It is not unusual
to find that for a given substance the passage from liquid to gas
is or is
not marked by a true phase transition depending on the
value of the amibient pressure, which could not happen if there
were an order parameter marking the difference.

It would appear that the partition function
$Z~=~{\rm Tr} e^{-H/T}$, being a sum of analytic terms, must
be an analytic function of the temperature -- and indeed,
in a finite volume,
it is.  However one is interested in the limit of large volumes, where the
sum in the trace may contain an infinite number of terms and analyticity
may be lost.  If the partition function is not analytic,
this must be due to some subtlety in taking the infinite volume limit.
One expects that such a subtlety must be associated with
the development of correlations which extend over an infinite range.
If the transition is second order, the correlation length must
go {\it continuously\/} to infinity at the transition point.  In particular,
sufficiently close the transition point the correlation length
will greatly exceed
any ``microphysical'' length associated with the problem,
\eg the lattice spacing (or, for QCD, $\Lambda^{-1}$).
These observations lead to the concepts of {\it scale invariance\/}
and {\it universality}, which together form the foundation of the modern
theory of second order phase transitions.

Scale invariance follows from the idea that non-trivial
correlations on length scales
much larger than the microscopic one can only depend on the ratio of
the length in question to one overall diverging correlation length.
(Generically there should be just one fundamental
diverging length, since there
is one parameter -- the temperature -- which must be tuned to reach the
critical point.  Another way of saying this is that all divergent
quantities should be given as functions of the reduced temperature
$t~=~|{T-T_c\over T_c}|$.)  Universality follows from the idea that
only modes with long-wavelength fluctuations in equilibrium
and their low-dimension interactions
among themselves are important for the singularities, and that
the only long-wavelength modes that exist are likely to be
those required by
the symmetries of the order parameter.
Together these principles tell us that in constructing models of
the singularities of thermodynamic functions near second order
phase transitions, it is necessary and sufficient to look for the
simplest scale-invariant models in the appropriate
dimension with the appropriate symmetry.

How does all this apply to QCD?  What are the available order
parameters?

For confinement, the only known order parameter is
the Polyakov loop [\pol ]
$$
P~=~ \langle \exp  \int i A_\tau d\tau \rangle
\eqn\aa
$$
where one is taking a thermal average, $A_\tau$ is the
imaginary-time component of the gauge field written as a
matrix in the {\it fundamental\/} representation, and the
trace is over one period $\beta~=~ T^{-1}$ of the imaginary
time.  $\ln P$
is basically the free energy associated with adding
a static color source with the quantum numbers of a quark.
It vanishes when quarks are confined.  However it
has no reason to vanish when quarks are not confined, and
indeed it does not vanish in perturbation theory.

The Polyakov loop
is not expected to
vanish when the color flux associated with
the color source can be screened.  Indeed the reason why
the energy cost of
inserting the color source in the confined phase is
infinite is because there is a non-trivial flux extending from the
source
that cannot terminate,
and this flux disrupts the vacuum wherever it threads, costing
a finite energy per unit length.  On the other hand if we are
considering a gauge theory with dynamical quarks, then the
quarks that are inevitably present at non-zero temperature
need only re-arrange themselves a bit to soak up the flux.  Strictly
speaking one then cannot distinguish confinement from screening.
There is no strict order parameter for confinement in the presence
of dynamical quarks at finite temperature (similar to the
case of ionization of an ordinary gas).

For
an $SU(N)$ gauge group, the Polyakov loop parameter transforms
non-trivially under the center of the gauge group, that is
$Z_N$.   When it vanishes the global
$Z_N$ symmetry under gauge transformations
which approach these central elements at infinity is unbroken, while
if it does not vanish this symmetry is broken.  (Note that
{\it local\/} gauge invariance is never broken.)

The universality class of the deconfinement transition for pure
glue $SU(N)$ is therefore described by a simple model with a
discrete global
symmetry group.  For $N=2$ it is the Ising model and for $N=3$
it is the
three-state Potts model.  These are well studied models.
The first of them has a second-order
transition in 3 dimensions, while the
second, for quite fundamental reasons, does not.  This led
Svetitzky and Yaffe [\sy ] to predict that pure glue $SU(2)$ would
probably have
a second-order, but pure glue $SU(3)$ must certainly have a
first-order, transition.  After some struggles, their prediction was
verified by direct numerical simulation [\nmglue ].

For the chiral transition with $N_f$ massless flavors, the
simplest order parameter one can imagine is given by an
$N_f\times N_f$ matrix $M$ of scalar fields, which parametrizes the
condensate
$\langle \bar q^i _L~ q_{jR} \rangle$.  This matrix is complex.
For $N_f~=~2$ one can require that $M$ is of the form
$\sigma~+ ~i\vec\pi \cdot \vec \tau$, with four independent real
components.  It turns out that $N_f~=~2$ is a very special case [\pw ].
It is in the universality class of a four-component magnet, a
model which has been extensively studied in the condensed matter
literature [\bnm ].
This model is known to have a second-order transition.
Thus there is a scale invariant candidate theory to describe the
singular behavior near a
second-order chiral symmetry restoration transition in 2 massless
flavor
QCD.  For more than two flavors the situation is very different.  The
more complicated models appropriate to this case have not been studied
so extensively.  However all the evidence,
both analytical [\pat ] and numerical [\gs ],
is that they do {\it not\/} support a second-order transition.  Thus
there is no candidate theory to describe the singular behavior that would
occur near a second-order chiral symmetry restoration in QCD with 3
or more massless flavors.  A probable interpretation [\pw ] is that
in these cases near
the transition fluctuations grow so large that they induce a first-order
transition.

These considerations suggest the possibility of a second-order transition
uniquely in the case $N_f~=~2$.  Remarkably, this pattern is consistent
with existing numerical evidence [\nmquark ].

When one has a second-order transition, it is possible to make
some precise predictions for the nature of the singularities near
the transition, using the powerful concepts of universality and
scale invariance as mentioned above.  Let me emphasize again that
these are {\it precise\/} predictions.  The most characteristic
results are predictiions of the critical exponents, which specify the
power dependence of various quantities such as the magnetization
(which in QCD language translates to the magnitude of the chiral
condensate)
the specific heat, or many others, on $t~=~|{T-T_c\over T_c}|$,
These predictions are similar in spirit to the famous predictions
for the behavior of moments of QCD structure functions as calculable
powers of logarithms of $Q^2$, but the calculations involved are
much more arduous.  The point is that whereas the ultraviolet fixed
point of QCD occurs at zero coupling, the infrared fixed point governing
the chiral phase transition occurs at a very large value of the coupling,
and extraordinary methods have to be used to extract quantitative results.
Fortunately the requisite work, involving calculating hundreds of
graphs in perturbation theory up to six loops,
using analytic methods to estimate the asymptotic behavior of
high orders of perturbation theory, and joining the two by sophisticated
resummation methods, was performed in a {\it tour de force\/} by
Baker, Meiron, and Nickel.

If we accept that for two massless quarks the transition is
second order while for three massless quarks it is first order, there
is an interesting question how one behavior goes over into the
other as the mass of the third quark varies.
A very pretty possibility is that as the third (strange) quark mass
is raised continuously from zero the discontinuities associated with
the first-order transition get smaller and smaller, eventually vanishing
at the so-called {\it tricritical point},
after which the transition becomes second-order.   The behavior in the the
immediate neighborhood of the tricritical point is governed by a
four-component asymptotically free massless $(\phi^2 )^3$ theory, which
features logarithmic corrections to mean field theory.
The tricritical point is characterized by a particular value $m_{\rm crit.}$
of the
strange quark mass.  Existing numerical evidence while very crude
suggests that the physical strange quark mass is larger than
$m_{\rm crit.}$.  However it is not impossible that the stange
quark mass
is quite close
to the critical value,
and of course in numerical experiments one can in principle
vary the masses to probe the suspected tricritical region.  It is
noteworthy that the specific heat, which generically has a cusp at the
second-order transition, develops an actual discontinuity at the tricritical
point.

Another direction in which the analysis may be extended is to
the consideration of time-dependent behavior near the critical point.
One can analyze how the transport equations change near the
critical transition.  The main qualitative phenomenon is critical
slowing: there is a diverging correlation time as well as a diverging
correlation length, as the system finds it difficult to relax
the very long-wavelength fluctuations that occur.
Thus transport coefficients generally acquire singularities
at the transition.  Dynamical critical
behavior probes more physics than the static behavior: models in the
same static universality class can have different dynamical critical
exponents; for example, ferromagnets and antiferromagnets are in the
same static universality class but different dynamic universality classes,
basically because the ferromagnetic order parameter, the magnetization,
becomes a conserved quantity in the long-wavelength limit
-- unlike the staggered magnetization -- which makes long-wavelength
fluctuations relax more slowly.  QCD with two massless quarks
appears to be in the dynamic universality class of the four-component
Heisenberg {\it antiferromaget}.   A major quantitative consequence of
the analysis is that the correlation time is predicted to scale as
the 3/2 power of the correlation length.  Note that if we translate this
behavior into a prediction for the form of real-time Green functions
we get funny cuts at zero frequency and wave vector, which is very
different
from a naive particle exchange model and of course from any naive
extrapolation of the behavior of weakly interacting plasma.

If we take the indications for a second-order phase transition with
two massless quarks and the effective ``massiveness'' of the
physical strange quark at face value, what are the consequences?
For numerical experiments, the we find ourselves in a happy situation.
There is a
wealth of detailed, quantitative predictions waiting to be tested.
For cosmology the situation is either dull or reassuring, depending on
your point of view.  The expansion of the universe is very slow
indeed on the relevant strong-interaction timescales, and equilibrium
should be closely maintained.  The non-zero masses of the up and down
quarks mean that there is really no phase transition at all.  Interesting
relics that might have occurred for a strongly first-order transition,
including gravity waves, inhomogeneous nucleosynthesis, and possible
production of exotic matter, do not occur.  For heavy ion collisions,
{\it if\/} thermal equilibrium is produced and maintained, the conclusion
is similar.  Even the interesting long-range correlations that might
have been expected to arise near a second order transitions, and to give
rise to interesting phenomenological signatures as I shall discuss in a
moment, are not quantitatively significant due to the finite pion masses
(that is, inverse correlation lengths) that are
not much smaller, and maybe not at all smaller, than the temperature
at the transition.   However that's a very big {\it if\/} as you know, and
it is interesting to consider another quite different idealization of
the physics.

\chapter{Quenching and Misaligned Patches}

If instead of cooling a bar of iron slowly through its
Curie point one suddenly plunges it into ice water, one is said
to have {\it quenched\/} the magnet.  It may be that it is not
inappropriate to model what occurs following a heavy ion collision,
as the hot plasma comes into contact with cold empty space outside,
as a quench.  Insofar as it is plausible to map the important
degrees of freedom for QCD into a magnet model, \ie\ if the
pions and the chiral condensate are mainly what we must
keep track of,
the analogy is quite close.

The dynamical evolution following a quench is quite different from
that for cooling through equilibrium.  One expects that broadly speaking
whereas near equilibrium the question is the typical
size of correlated regions,
which individual grow, shrink, come into being and pass away, following
a quench the system is racing to the ordered state, the dynamics is
unidirectional, and the question is how the domains evolve (grow) in time.

There is a substantial condensed matter literature on problems of this
type, including very recent work [\bm ].
Also related problems arise in
the cosmological models (``texture models'') where the fluctuations
responsible for triggering the formation of
structure in the universe are ascribed to the formation, growth, and
eventual collapse of domains during a cosmic phase transition [\txtr ].
The
QCD problem has its own special features, however, and requires a fresh
analysis.  The main new features that are central to the QCD problem are
the fact that the symmetry is intrinsically broken, and that one considers
hyperbolic rather than diffusive equations.  In many condensed matter
problems diffusive equations for the dynamical variables of interest are
appropriate, because these variables are in contact with other degrees
of freedom (\eg\ phonons ) whose effect is modelled by appropriate diffusive
transport equations.  However in the QCD problem, and probably also in
some condensed matter problems, one should use the true microscopic equations
which of course are hyperbolic.

The fact that the symmetry is intrinsically broken means that
every region of space is heading toward the same ground configuration,
and interest focuses on the nature of the approach.
Do different regions of space relax independently to the ground configuration
(small oscillations around the $\sigma$ direction), or do large aligned
regions form and then relax coherently?   The latter possibility could have
dramatic consequences, because it would mean that large regions of
space have a misaligned condensate [\misal ],
a classical field configuration that
might be expected to emit classical pion radiation  -- to ``lase'' pions
-- as it relaxes.

To model a QCD quench we take a representative configuration
of the $\sigma$ and $\pi$ fields from the
ensemble weighted by the free energy at the pre-quench temperature
$T$, and then evolve it according to the zero temperature equations of
motion.   We find that for broad ranges of the parameters
large coherent structures do form following a quench.  We believe that
the fundamental mechanism underlying this behavior is the following.
The (mass)$^2$ of a Nambu-Goldstone field is zero in the true ground
state, due to a cancellation between an intrinsically negative
bare (mass)$^2$ and a positive contribution from the interaction with
the condensate:
$$
m^2~=~ - \mu^2 + 2\lambda v^2 ~=~ 0 ~({\rm ground state})
\eqn\masseq
$$
However in a quench the vacuum expectation value $v^2 ~=~
\langle \sigma \rangle ^2 $ starts centered around zero, not its
ground state value $\mu^2/2\lambda$.  Thus $m^2$ can easily be
negative, and according the the dispersion relation
$$
\omega^2 ~=~ k^2 + \mu^2
\eqn\disprel
$$
modes with sufficiently small spatial frequencies
will grow -- and the
longer the wavelength, the more the growth.
Though the real situation
(that is, our idealized model of a quench)
is more complicated in various ways, including
the non-linearity of the equations, the phenomenon suggested by
this simple picture does occur roughly as expected.

Taking a realistic intrinsic symmetry breaking into account does
not destroy the phenomenon, because the intrinsic (mass)$^2$ of the
pion is substantially less than $\mu^2$.  (The ratio of the two
is basically the square of the ratio of pion and sigma masses; see
section 4 below).

Thus there seems to be a good chance that large regions of
misaligned vacuum might form.
The radiation from such a region as it relaxes
will be quite structured and
unusual.  The 4-component classical field $(\sigma , \vec \pi )$
will oscillate in the plane determined by sigma and some definite
direction of the vector $\vec \pi$.  The radiation is a coherent
configuration of pions {\it in that direction}.  These radiated
pion clumps
should have small relative momentum, and each one will have a
fixed ratio of charged to neutral pions.  The probability distribution
for a given charge ratio is
$$
{\rm Prob.} ({\cal R } ) ~=~ {1\over 2} {\cal R}^{-{1\over 2}}
\eqn\piratio
$$
where
%
$$
{\cal R} ~\equiv~ {\pi^0 \over \pi^0 + \pi^+ + \pi^-} ~.
\eqn\defr
$$
This distribution is, of course, highly non-Gaussian.  For example
the probability of having less than 1\% neutral pions is
10\% !

\chapter{Equilibrium Phase Transition One Dimension Down}

It would be interesting to test
the logic of the argument leading to the expectation of a possible
second order deconfinement
phase transition for color $SU(2)$ pure glue theory and chiral
symmetry restoration phase transition for
two flavors of massless quarks, but definitely first order for
deconfinement in color $SU(3)$ pure glue theory or chiral symmetry
restoration with three or more flavors, in other examples.
A simple possibility lies near at hand: one can study the corresponding
questions in one fewer spatial dimension.  While the
direct relevance of
such studies
to any achievable laboratory situation is doubtful, they have
the advantage
of being relatively easy to access by numerical experiments.
Also as we shall see some rather striking situations might arise.

Following the same arguments as before, we relate the possible
second order transitions of 2+1 dimensional gauge theories
to the existence of 2 dimensional scale invariant theories with
the same symmetries.

The study of 2 dimensional scale invariant theories has been a
large thriving activity in recent years [\cft ].
In two dimensions scale
invariance leads to a much larger and more powerful symmetry group
of conformal symmetry transformations, and use this symmetry to
carry the analysis of scale or conformal invariant
theories a long way.
In favorable cases critical exponents can be calculated exactly
and there are tractable algorithms for calculating just about
any correlation function of interest.
Thus a question of interest becomes, which conformal field
theories correspond to which gauge theories at their critical points.

It seems quite plausible that the $SU(N)$
pure glue theories have deconfinement phase transitions associated with
the $Z_N$ central symmetries discussed above.  For $N=2$ one has
the symmetry class of the Ising model, for $N=3$ the three-state Potts
model, and so forth.
Whereas in 3 dimensions the phase transition was second order for
$N=2$ but necessarily first order for $N=3$, here they can both
be second order.  As I said there is a quite a full undertanding of
the correlation functions of these models, though it may not be
entirely trivial to set up the dictionary translating these
results into
gauge theory language.

The issue of chiral  symmetry breaking in 2+1 dimensions is
complicated by two factors, which make the situation very
different from what one has in 3+1 dimensions, where one has
breakdown of a continuous chiral symmetry.
First, there is no chirality in 2+1 dimensions; and second, there
can be no spontaneous breaking of a continuous symmetry. Let me
recall these facts for you.

In 2+1 dimensions we can use (essentially) the ordinary
two by two Pauli matrices as gamma matrices, say
$(\gamma_0, \gamma_1, \gamma_2 )~=~ (\sigma_2, i\sigma_1, i\sigma_3)$.
These are chosen in such a way that they are all imaginary, so that the
Dirac equation can be solved with real (Majorana) spinors.  However we
shall focus on fermions with the quantum numbers of quarks, which
are necessarily complex anyway.  More important for present
purposes is the fat that the product
$\gamma_0 \gamma_1  \gamma_2 $ reduces to a pure number.
This implies
that it is impossible to make a chiral projection.
Another fact that will be important to us is
that the parity operation of reflection in one axis
acts as (say) $\gamma_0 \gamma_2$ on the spinors, so that the
operator $\bar \psi \psi$ is {\it odd\/} under parity (and time
reversal).

There is
no spontaneous breaking of continuous symmetries in 2 dimensional
quantum field theory
at zero temperature (Coleman theorem, [\cole ])  --
or for
2+1 dimensional theories at non-zero temperature
(Mermin-Wagner theorem, [\merm ]).  These are very nearly the same theorem,
since the
zero-frequency components of the putative Nambu-Goldstone fields
have the infrared singularities (long wavelength fluctuations)
of the 2 dimensional theory.  These fluctuations are
actually divergent at long distances;
their divergence undoes the hypothesized
symmetry breaking or long-range order.  This shows
the internal
inconsistency of that hypothesis that the continuous
symmetry breaks,
leading to the theorems.

Still there is something to investigate.
The theory with $f$ flavors of massless quarks has a $U(f)$ flavor
symmetry and also a parity symmetry.  As we have seen even a
common mass terms, or a
quark-antiquark condensate without any preferred direction in flavor
space, breaks parity.  Thus if we assume that a
flavor singlet condensate develops
at zero temperature, but goes away at high temperature, then there
is the possibility of a phase transition with $Z_2$ symmetry,
conrresponding
to parity restoration.  If this phase transition is second order, then
it could be mapped onto the Ising transition.



I think it would be quite interesting to carry out
appropriate numerical
experiments on lower-dimensional QCD
at finite temperature, to check whether second-order transitions
with the predicted exponenets in fact occur.
Hansson and Zahed [\hz ] have also emphasized that the
high temperature behavior of the 2+1 dimensional gauge theory appears
to be tractable theoretically and to form a good testing ground for
analytical methods which can also be applied
to the more difficult 3+1 case.

\bigskip

\bigskip

\chapter{Energy Budget for Chirality and Deconfinement}

It is an interesting exercise to estimate the energy
locked up in the chiral vacuum, and to compare it
with the energy of the glue near the
phase transition.

To estimate of the energy density in the chiral vacuum,
begin with the sigma model potential
$$
V(\sigma ) ~=~ -\mu^2 \sigma^2 ~+~ \lambda \sigma^4~.
\eqn\sigmapot
$$
Elementary calculations lead to the vacuum expectation value
$\langle \sigma \rangle ~=~ \sqrt{ \mu^2 / 2\lambda } $, the
sigma mass $m_\sigma ~=~ 2\mu $ and the energy density
${\cal E } ~=~ - \mu^4/ 4\lambda$ at the minimum, where the
symmetric vacuum energy density is normalized to zero.  Identifying
$\langle \sigma \rangle ~=~ f_\pi $ we arrive at the estimate
$$
{\cal E} ~=~ - {f\pi^2 m_{\sigma}^2 \over 8}
\eqn\sigenerg
$$
of the condensation energy in terms of observables.
Inserting experimental values, one finds
$$
|{\cal E}| ~\approx~ 5 \times (100 ~{\rm Mev})^4~.
\eqn\signume
$$
This estimate, based on classical reasoning and on taking the
broad observed
$\sigma$ resonance as the embodiment of the field in the
sigma model, is certainly crude.  However it may not be inappropriate
to use it to draw one tentative qualitative conclusion, as follows.

The energy density for an ideal gas of $SU(N)$ color gluons
at temperature $T$ is
$$
{\cal E_{\rm glue} } ~=~ (N^2 -1){\pi^2\over 15} T^4
{}~\approx ~ 4.8 \times T^4 ~({\rm for}~ N=3) ~.
\eqn\gluee
$$
There is also a contribution from $N_f$ species of
massless quarks,
with the prefactor ${7\over 4} N N_f$ replacing $N^2 -1$ in
\gluee .   For $N=3$ and two light flavors, this quark contribution
is just a bit larger than \gluee .  The simple
qualitative point I want to stress
is that the sorts of energy densities associated with the
chiral condensation are, for $T~\sim~ 100 ~{\rm Mev}$ and
for the stated values of $N$ and $N_f$, roughly
comparable to the energy densities of the entire quark-gluon
plasma.  This comparison is important in thinking about the
question whether the singular parts of the thermodynamic functions,
which are the universal quantities we can address theoretically,
are quantitatively important in the relative to the thermodynamic
functions themselves.  For example is the predicted cusp in specific
heat a major feature or just a tiny dimple on an otherwise smooth curve?
This is a non-universal question, whose answer will depend on details
of the microscopic theory including the absolute value of
the transition temperature (which is very much conditioned on such
details).  The estimates above  suggest that if
$T~\approx~ 100 ~{\rm Mev}$, as is perhaps suggested by the numerical
work, then the energy controlled by the value of the order parameter
is not much less than the total energy, so that the singular parts
should may not be overly difficult to discern.  Existing evidence suggests
that $T$ is a little larger than this, but on the other hand probably one
should
not expect the gluon plasma to acquire its full (non-interacting) energy
immediately following the transition.
In any case, it is important to  emphasize that the foolproof way to look
for manifestations of critical singularities is to study the behavior of
the order parameters and closely related quantities, which is completely
dictated by the universal theory.

The issue of the energy difference is also important in assessing
the appropriateness of the quench model as an
idealization.  The essence of the
quench phenomenon is that the modes associated with spontaneous
symmetry breaking are slower to relax to equilibrium
than the generality of
modes.  Thus when energy is suddenly drained from the
quenched system these modes
are trapped in the pre-existing configuration, but at a low temperature.
Clearly the most favorable case is when the energy associated with the
symmetry-breaking modes is commensurate with the total energy, so that
we are not making absurd demands on the cooling process.  Expansion
of the plasma provides an
excellent cooling mechanism, but we
should not require miracles of it.

\ack{As mentioned before, the first two sections report joint work
with Krishna Rajagopal, who also made some helpful remarks on the
other sections.  I am also grateful to Chetan Nayak and to
Bert Halperin for important insights regarding the 2+1 dimensional theory.}

\endpage

\refout

\end

\chapter{Question of Charge Separation at a Quench}

The striking prediction \piratio for the probability distribution
of neutral relative to charged pions naturally leads to the question
whether there are similary unusual fluctuations in the charge itself,
\ie in the ratio of $\psi^+$ to $\pi^-$ mesons.  Let me first warn you
against an elementary confusion: the charged pions are {\it not\/}
related by an isospin transformation to the neutral one.  Rather it is
the real fields $\pi^1~=~{1\over \sqrt 2} (\pi^+ + \pi^-)$
and $\pi^2~=~{1\over \sqrt 2 i} (\pi^+ - \pi^- )$ or linear
combinations of them with real coefficients into which $\pi^0$ rotates.
Thus one cannot conclude simply from isospin invariance that a formula
similar to \piratio , but with a charged pion field in the numerator,
is valid.

In fact it is not.  The correct equation is instead
$$
{\rm Prob.}  ({\cal R'} ) ~=~ (1- 2{\cal R'} )^{-1}
\eqn\charrat
$$
where
$$
{\cal R'}~=~ {\pi^+ \over \pi^+ + \pi^- + \pi^0 }
{}~=~  {\pi^+ \over \pi^+ + \pi^- + \pi^0 } ~.
\eqn\defrat
$$
The equality
of the last two terms in \defrat\ should hold to a good
approximation event by event, and ${\cal R'} \leq 1/2$.
This is just an obscure
way of saying that the charged pions do not
separate.

The point is that the charge operator
$$
j_0~=~ \pi_2 {\partial \over \partial t} \pi_1
- \pi_1 {\partial \over \partial t} \pi_2
\eqn\chrgop
$$
measures rotary motion in the 1-2 plane.  However our amplification
mechanism gives a direct outward kick to the field, and does not
induce rotary motion.  This heuristic argument is borne out in the
simulations.

Notice that the question of whether charge separation occurs is
a dynamical one, and we have discussed it in a straightforward dynamical
way for the quench mechanism of generating misaligned vacuum.
In earlier work on this question (based on the intuitive ``baked Alaska''
model) Kowalski and Taylor
imposed isospin symmetry by hand
in order to avoid the possibility of charge separation,
which they consider physically implausible.
The coherent states we reach
are {\it not\/} isospin singlets, and we see no reason to
impose that condition; nevertheless the intuition of Kowalski and Taylor
is vindicated, and there is no charge separation.

\end